\documentclass[reprint,
superscriptaddress,
 amsmath,amssymb,
 aps,
 prl,
]{revtex4-1}

\newcommand*\chem[1]{\ensuremath{\mathrm{#1}}}

\usepackage{graphicx}
\usepackage{dcolumn}
\usepackage{bm,xcolor,siunitx}
\usepackage{amsmath,amssymb}
\usepackage{subfigure}
\usepackage{mdframed}
\usepackage{hyperref}
\usepackage{bm,placeins}

\begin{document}

\title{Universal Model for the Turn-on Dynamics of Superconducting Nanowire Single-Photon Detectors}

\author{Kathryn L. Nicolich}
\affiliation{Department of Physics, The Ohio State University, Columbus, OH 43210, USA}
\author{Clinton Cahall}
\affiliation{Department of Electrical and Computer Engineering, Duke University, Durham, NC 27708, USA}
\author{Nurul T. Islam}
\affiliation{Department of Physics, The Ohio State University, Columbus, OH 43210, USA}
\author{Gregory P. Lafyatis}
\affiliation{Department of Physics, The Ohio State University, Columbus, OH 43210, USA}
\author{Jungsang Kim}
\affiliation{Department of Electrical and Computer Engineering, Duke University, Durham, NC 27708, USA}
\affiliation{IonQ, Inc., College Park, MD 20740, USA}
\author{Aaron J. Miller}
\affiliation{Quantum Opus LLC, Novi, MI 48375, USA}
\author{Daniel J. Gauthier}
\affiliation{Department of Physics, The Ohio State University, Columbus, OH 43210, USA}

\date{\today}

\begin{abstract}
We describe an electrothermal model for the turn-on dynamics of superconducting nanowire single-photon detectors (SNSPDs).  By extracting a scaling law from a well-known electrothermal model of SNSPDs, we show that the rise-time of the readout signal encodes the photon number as well as the length of the nanowire with scaling $t_\text{rise}\propto \sqrt{\ell/n}$.  We show that these results hold regardless of the exact form of the thermal effects.  This explains how SNSPDs have inherent photon-number resolving capability. We experimentally verify the photon number dependence by collecting waveforms for different photon number, rescaling them according to our predicted relation, and performing statistical analysis that shows that there is no statistical significance between the rescaled curves. Additionally, we use our predicted dependence of rise time on detector length to provide further insight to previous theoretical work by other authors.  By assuming a specific thermal model, we predict that rise time will scale with bias current, $t_\text{rise}\propto \sqrt{1/I_b}$. We fit this model to experimental data and find that $t_\text{rise}\propto 1/(n^{0.52 \pm 0.03} ~I_b^{0.63 \pm 0.02})$, which suggests further work is needed to better understand the bias current dependence.  This work gives new insights into the non-equilibrium dynamics of thin superconducting films exposed to electromagnetic radiation.
\end{abstract}

\maketitle


Superconducting nanowire single-photon detectors (SNSPDs) are widely used in quantum optics and quantum information science because of their  high efficiency over a wide range of wavelengths, fast reset times, low timing jitter, low dark count rates, and typical lack of afterpulsing \cite{dauler2014review,hadfieldbook2015}.  Despite widespread use of these detectors, their dynamics are still not fully understood.

A qualitative picture of SNSPD operation involves one or more photons absorbed by the device simultaneously that create resistive regions in the nanowire, known as hot-spots, which divert current out of the detector and into the readout circuit, constituting a detection event.  A complete, quantitative model requires knowledge of the spatial-temporal dynamics of the non-equilibrium distribution of quasi-particles in the superconductor and its interaction with the readout electronics.  The understanding of the microscopic details of these devices is rapidly advancing \cite{engel}; however, electrothermal models of SNSPDs frequently have more parameters than measurable constraints \cite{yang2007,kerman2009electrothermal}. Hence, similar behaviors may be fit using different sets of model parameters, thus potentially obscuring the physical principles underlying SNSPD behavior.
  
Here, we describe a simple SNSPD model that captures the essential physics of the link between hot-spot growth and features of the rising edge of the electrical readout pulse.  The model identifies a `universal curve' for the electrical signal.  Converting this model to physical units requires using only two scale parameters, which can be determined experimentally, and gives a simple relation between the microscopic SNSPD parameters and the readout signal.  Using scaling relations derived from our model equations, we explain the recent demonstration of multi-photon resolution in a conventional SNSPD \cite{cahall2017heckyes}.  Further, we predict the scaling of the turn-on time with nanowire length $\ell$ and find results consistent with a previous prediction that relies on a more complex model.  Finally, by refining our model in a way suggested by the work of Kerman \textit{et al.} \cite{kerman2009electrothermal}, we predict the specific shape of the rising edge of a pulse and the dependence of the rise time on the bias current $I_b$ of the detector. We claim that our theoretical predictions hold regardless of detector material, and we compare to experiments performed with a detector made from a proprietary amorphous superconductor in a similar class as WSi and MoSi, as well as results from the literature on NbN. We find good agreement between the model predictions and experimental measurements.

\begin{figure}[h]
\centering
\includegraphics[width=3.375in]{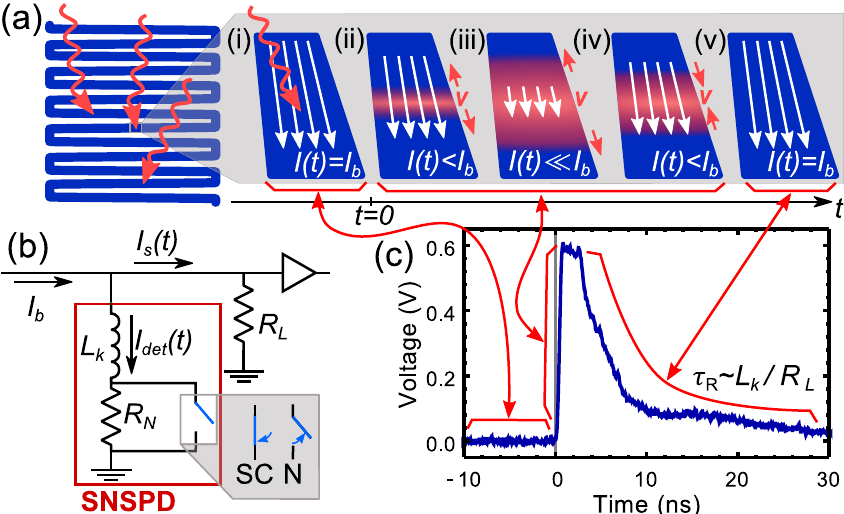}
\caption{(a) Depiction of an SNSPD meander with incident photons. (a-i) Illustration of a region of the SNSPD near an incident and absorbed photon. (a-ii:v) The progression of hot-spot formation, growth, and decay, which is described using the front velocity $v$. (b) Lumped-element circuit diagram of an SNSPD and readout. (c) An example SNSPD readout pulse. The vertical gray line marks $t=0$.}
\label{snspd}
\end{figure}

A typical high-detection-efficiency SNSPD consists of a thin ($\sim$~5--10-nm thick) and narrow (width $w\sim$ 100 nm) superconducting film shaped in a meander that matches the optical field mode of the photon source as illustrated in Fig.~\ref{snspd}(a).  Electrically, we treat the SNSPD using lumped circuit elements coupled to a readout circuit, shown in Fig. \ref{snspd}(b).  The detector bias current splits between two pathways to ground: current $I_\text{det}$ passing through the SNSPD and the signal current $I_s$ passing through the readout circuit load resistor $R_L$.  The SNSPD is treated as a kinetic inductance $L_k$,  connected in series with a parallel combination of a time-dependent resistance, $R_N(t)$ and a switch. The resistive part of the nanowire due to photon detection is represented with $R_N(t)$, and the switch in the closed state describes the entire device being superconductive.  In Ginzburg-Landau theory, $L_k$ depends on $I_\text{det}$.  The variation is essentially constant when the current is below $\approx$ 90$\%$ of the depairing current and decreases rapidly as $I_\text{det}$ approaches the depairing current from below \cite{clemandkogan}. For most present day SNSPDs, $I_\text{det}$ is well enough below the depairing current that taking $L_k$ constant is a good approximation and we assume that here. This approximation is discussed further in Appendix A. In the absence of photons, the nanowire is in thermal equilibrium with the substrate at temperature $T_o<T_c$, where $T_c$ is the superconducting critical temperature at zero bias, and the switch is closed ($I_\text{det}=I_b$, $I_s=0$).

A detection event begins with the absorption of a photon by the nanowire with an energy much greater than the superconducting gap energy (Figs.~\ref{snspd}(a-i)), giving rise to a hot-spot. We use the so-called hot-spot mechanism to describe detector operation. While there continues to be some uncertainty of the appropriateness of this model during the very short time during the initiation of a detection event \cite{bulaevskii2012,renema2014,casaburi2015,engel1015,gaudio2016}, it predicts many experimental observations. For the purposes of this paper, we consider only the dynamics after the hot-spot has already grown across the entire cross-sectional area of the nanowire \cite{semenov2001}.  Note that throughout this work we use the term hot-spot to refer to the normal region as it grows and shrinks due to Joule heating, in contrast with the localized resistive region present immediately following photon absorption before Joule heating takes place. The hot-spot spreads rapidly due to quasiparticle diffusion and extends across the width of the nanowire (Fig.~\ref{snspd}(a-ii)), giving rise to a resistive (normal) wire segment with resistance $R_\text{hs}$. In the lumped-element circuit model, the switch is now open.

The hot-spot continues to grow along the length of the wire due to Joule heating from current $I_\text{det}$ passing through the normal region and this growth is characterized by a phase front velocity $v=v(I_\text{det})$ at its boundaries (Fig.~\ref{snspd}(a-iii:iv)).  This treatment is widely used to describe SNSPD dynamics and is appropriate as long as the thermal healing length $L_H$ is much smaller than the length of the hot-spot. For NbN based detectors, $L_H=$20 nm, therefore this approximation is valid for all but the very early initial stages of hot-spot development  \cite{yang2007,kerman2009electrothermal}. As the resistive region grows, more current is shunted into the readout circuit, resulting in the rising edge of the electrical pulse shown in Fig.~\ref{snspd}(c).  As more current is shunted out of the detector, the hot-spot growth slows and stalls at a steady-state current $I_\text{ss}$.  (Some authors call this the retrapping current \cite{hazra2010,smirnov2016}).

Here, we focus on the time interval during hot-spot growth and decay up until hot-spot collapse, which corresponds to the rising edge of the electrical pulse.  This sequence of events takes place on a short time scale, typically less than 1 ns.  Previously, the rising edge of the readout signal has been described as an exponential growth with time constant $L_k/R_\text{hs}$ \cite{natarajan2012review}, where $R_\text{hs}\sim$ 10$^3$ $\Omega$ is constant during the presence of the hot-spot. However, as we show below, this is not an accurate picture and leads to incorrect conclusions about the detector turn-on dynamics.

Shortly after its growth stalls, the hot-spot collapses as cooling to the substrate dominates over Joule heating, as shown in Fig.~\ref{snspd}(a-v). Once the nanowire returns to the superconducting state, the switch in the lumped-element circuit model is closed and the readout pulse begins to decay exponentially on a time scale governed by the time it takes for the current to return to the inductor (Eq.~\ref{elecEQ} with $R_\text{hs}=0$), which has a 1/$e$ recovery time of $\tau_r=L_k/R_L$. Note during the decay of the pulse, both $L_k$ and $R_L$ are constant. For high-speed readout circuits, $R_L$ is often 50 $\Omega$, so that $\tau_r$ is tens of nanoseconds - much longer than the rise time of $I_s$.

The dynamics of the coupled system during a detection event are governed by the interplay between the superconducting nanowire and the readout circuit, which we describe using an electrothermal model given by  
\begin{eqnarray}
\label{genericthermEQ}\frac{dR_\text{hs}}{dt} &=& 2\frac{R_\text{max} }{\ell} v(I_\text{det}) \hspace{.2in}  R_\text{hs} \geq 0, \\
L_k\frac{dI_\text{det}}{dt}+n R_\text{hs}I_\text{det} &=& (I_b-I_\text{det}) R_L, \label{elecEQ} 
\end{eqnarray}
which are valid immediately after the initiation of a detection event, with initial conditions $R_\text{hs}(0)=0$ and $I_\text{det}(0)=I_b$.  Here, $R_\text{max}$ is the resistance of the nanowire when its total length $\ell$ is in the normal state, which occurs when $T>T_c$ at $I_b=0$ or $I_b \gg I_\text{sw}$ at $T=0$, where $I_\text{sw}$ is the switching current. Thus $R_\text{max} / \ell$ is the resistance per unit length of the nanowire and Eq.~\ref{genericthermEQ} expresses the above description of hot-spot growth. The factor of two accounts for the two fronts that bound a hot-spot. Physically, $R_\text{hs}>0$, but the system of equations allows for unphysical solutions arising from the superconductivity transition of the hot-spot.  Keeping the solution in the physical domain requires setting $R_\text{hs}=0$ when $R_\text{hs}$ falls below zero in solving the system of equations. Here we assume that $v(I_\text{det})$ only depends on the present value of $I_\text{det}$ and not on the size or history of the hot-spot. The physics of a propagating superconducting-normal boundary, traveling at a phase velocity $v(I_\text{det})$, for long, narrow superconductors has been studied extensively because of its importance in understanding quenching of superconducting magnets \cite{gurevich1987}.  In the latter half of this work, we use a particular form of $v(I_\text{det})$ well suited to most SNSPDs.

Equation~\ref{elecEQ} is Kirchhoff's voltage law for the SNSPD-load resistor loop. We allow for the possibility that $n$ photons are absorbed by the film simultaneously as might happen when illuminating the nanowire with a short-duration multi-photon wavepacket.  We assume that each absorbed photon generates a hot-spot with identical behavior and that they do not overlap spatially. This is appropriate for a small number of hot-spots with typical stalled hot-spot maximum lengths ($\sim$1 $\mu$m) and typical  values of $\ell$ ($\sim 500\ \mu$m) \cite{yang2007}.  The total nanowire resistance is then given by $R_N(t)=nR_\text{hs}(t)$.  While the hot-spots do not interact directly in our model, they are coupled indirectly through the electrothermal interaction.

To explore the effects of various parameters on detector rise times, we rescale Eqs.~\ref{genericthermEQ} and \ref{elecEQ}.  There are two timescales inherent in this system of equations, $\tau_r$ and $t_{ch}$; the latter of which is the characteristic timescale for the dynamics related to the rising edge of the readout pulse.  Because we are interested in studying this regime, we rescale Eqs.~\ref{genericthermEQ} and \ref{elecEQ} taking $t_{ch}$ to be the dominant timescale.  We introduce dimensionless quantities $\tilde{t}=t/t_\text{ch}$, $\tilde{\tau}_r=\tau_r/t_\text{ch}$, $\tilde{R}(\tilde{t})=R_\text{hs}(t)/R_\text{ch}$, $\tilde{R}_\text{max}=R_\text{max}/R_\text{ch}$, $\tilde{I}(\tilde{t})=I_\text{det}(t)/I_b$, and $\tilde{v}(\tilde{I})=v(I_\text{det})/v_b$, where $v_b=v(I_b)$ is the phase front velocity at the start of hot-spot formation.  Here, the characteristic resistance $R_{ch}$ represents the maximum resistance attained by a single hot-spot and it and $t_{ch}$ are given by
\begin{eqnarray}
R_\text{ch}&=&\sqrt{\frac{4 v_b R_\text{max} L_k }{n \ell}},
\label{Rch} \\
t_\text{ch}&=&\frac{2 L_k}{n R_\text{ch}}=\sqrt{\frac{\ell L_k}{n v_b R_\text{max}}},
\label{tch}
\end{eqnarray}
respectively. Substituting these expressions into Eqs.~\ref{genericthermEQ} and \ref{elecEQ} results in the dimensionless set of equations
\begin{eqnarray}
\frac{d\tilde{R}}{d\tilde{t}}&=&\tilde{v}(\tilde{I}) \hspace{.3in} \tilde{R} \geq 0,
\label{nearuni2}\\
\frac{d\tilde{I}}{d\tilde{t}}+2\tilde{R}\tilde{I}&=&\frac{1}{\tilde{\tau}_r}(1-\tilde{I}).\label{nearuni1}
 \end{eqnarray}
In the rescaling Eqs.$~\ref{Rch}$ and$~\ref{tch}$, the dependence on photon number $n$ was the unique choice for eliminating the explicit $n$ dependence in Eqs.~\ref{genericthermEQ} and~\ref{elecEQ}. There is still a ``hidden'' dependence in that systems with different numbers of photons will have different values for $ \tilde{\tau}_r $. 

For typical SNSPDs, the turn-on dynamics are much faster than the recovery time $\tau_r$, such that $\tilde{\tau}_r \gg 1$ ($n R_{ch} \gg 2 R_L$), and it is possible to neglect the right-hand-side of Eq.~\ref{nearuni1}.  Assuming we operate in this domain, the coupled equations become
\begin{eqnarray}
\frac{d\tilde{R}}{d\tilde{t}}&=&\tilde{v}(\tilde{I})\hspace{.3in} \tilde{R} \geq 0,  \hspace{.3in} 
\label{nearuni2f} \\\frac{d\tilde{I}}{d\tilde{t}}+2\tilde{R}\tilde{I}&=&0 ,\label{nearuni1f}
\end{eqnarray}
with initial conditions $\tilde{I}(\tilde{t}=0)=1$ and $\tilde{R}(\tilde{t}=0)=0$. $\tilde{v}(\tilde{I})$ plays the role of a driving term for the system of equations. We first consider the nanowire operating at a specific bias current $I_b$. The phase front velocity is taken to be some (perhaps unknown) function of the detector current $v(I)$. However, for all cases having the same bias current $I_b$, the scaled velocity function $\tilde{v}(\tilde{I})$ is identical -- these cases are all described by exactly the same scaled equations above and hence have exactly the same scaled solutions for  $\tilde{I}(\tilde{t})$ and $\tilde{R}(\tilde{t})$. In particular, the relative values for the characteristic time-scale for the turn-on dynamics $t_\text{ch}$ are independent of the specific form of $\tilde{v}$.

Therefore, for a detector operating at a specified $I_b$, we can directly relate the signal for a multi-photon event to that of a single-photon event by explicitly expressing the solution of Eqs. \ref{nearuni2f} and \ref{nearuni1f} for the scaled current in terms of the actual, physical quantities for the two cases:
\begin{equation}
I_{1}(t)=I_{n}(t/\sqrt{n}).
\label{unicurve}
\end{equation}
Thus, the absorbed photon number $n$ is encoded within the turn-on dynamics, which can be revealed by measuring the 10\% - 90\% signal rise time or the  maximum value of $dI_{s,n}/dt \propto \sqrt{n}~dI_{s,1}/dt$.  This explains the recent experimental findings of Cahall \textit{et al.} \cite{cahall2017heckyes}.

To compare our theoretical predictions to experimental observations, we collect waveforms from a proprietary amorphous device for optical wavepackets with $n=$1, 2, and 3 photons in the same manner as in Ref.~\cite{cahall2017heckyes}, using a higher-bandwidth read-out circuit to minimize signal distortion, and rescale them using the same principle as Eq.~\ref{unicurve} (see Appendices B and C for experimental details).  In Fig.~\ref{n123}(a), we show the rising edges of the waveforms collected at the readout.  Figure~\ref{n123}(b) shows the waveforms after rescaling by $\sqrt{n}$, where it is seen that the waveforms appear to fall on a single curve.  This phenomena is more apparent when examining the derivatives of the waveforms, which are shown in Fig.~\ref{n123}(c) and rescaled in (d).  The difference at $t=0$ in the derivative curves is statistically significant.  In contrast, there is no statistical significance between the rescaled derivative curves at $t=0$ (see Appendix D for details on the statistical analysis).  Therefore, we claim that rescaling the traces by $\sqrt{n}$ reveals a universal curve independent of photon number.  Ringing from our amplifier distorts this effect somewhat beyond $t_N=0.4$ ns in the rescaled curves.

\begin{figure}[h]
\centering
\includegraphics[width=3.375in]{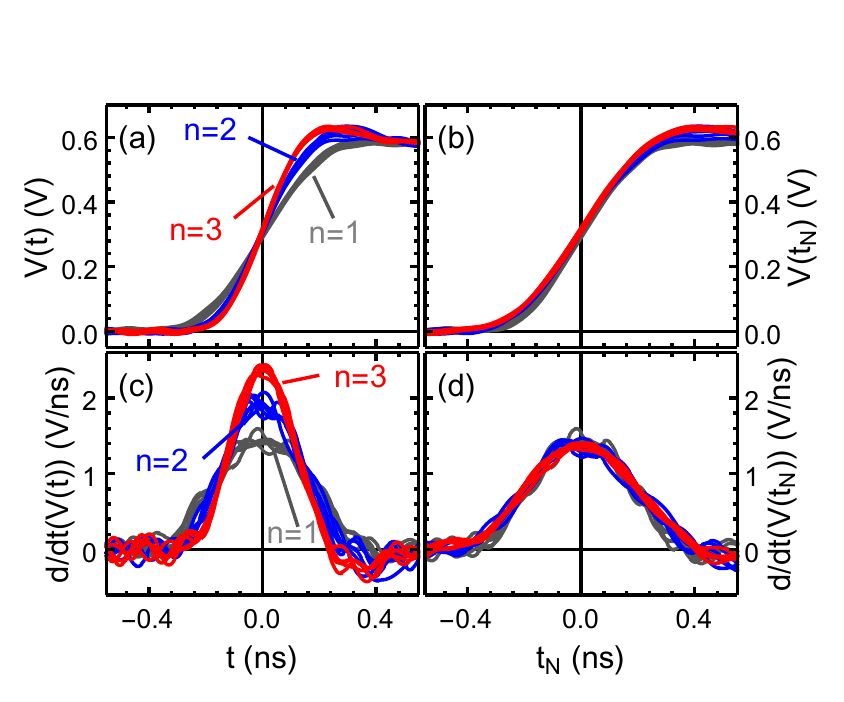}
\caption{(a) Rising edges of $n=$1, 2, 3 traces (gray, blue, red) and (b) with $t_N=t/\sqrt{n}$, as well as their derivatives with respect to time (c) and (d).}
\label{n123}
\end{figure}
Recent work by Smirnov \textit{et al.} provides a second test of our model \cite{smirnov2016}.  They modeled and experimentally studied the effect of detector length on single-photon pulse rise times for SNSPDs.  They present a two-temperature (quasiparticle/phonon distributions) model described by five coupled differential equations, which was numerically solved to predict readout signal rise times.   They tested their predictions by comparing signals from detectors with different $\ell$.  The authors argue, qualitatively, that the dependence of rise time on $\ell$ should be nonlinear for $\ell>20\ \mu$m and  their model equations confirm this.  While we cannot make absolute estimates of detector rise times, we can make a stronger claim of the nature of the nonlinear dependence on $\ell$. Detectors differing only in length (and operated at the same bias current) will all be described in our approach by Eqs.~\ref{nearuni2f} and~\ref{nearuni1f}. In Eq.~\ref{tch}, $L_k$ and $R_\text{max}$ are proportional to $\ell$.  Therefore, we predict $t_\text{ch}$ and hence  detector rise time, should scale as $\sqrt{\ell}$.  We fit their model calculations to $g+h \ell^{1/2}$, allowing $g$ and $h$ to be fit parameters, and find $g=15.1 \pm 0.8~\text{ps}$ and $h=15.01 \pm 0.04 ~\text{ps}/\chem{\mu m^{1/2}}$, with a reduced chi-square statistic of $\sim 1$.  (See Appendix E for more information.)  Our prediction for the scaling of the rise time on $\ell$ and $n$ does not require a solution to the differential Eqs.~\ref{genericthermEQ} and~\ref{elecEQ}, but is based only on the rescaling of the equations in physical units, Eqs.~\ref{Rch} and~\ref{tch}. Thus, we conclude that these scaling results are due to the thermoelectric coupling and do not depend on the microscopic physics that defines the functional dependence of $v$ on $I_\text{det}$.

\begin{figure}[h]
\includegraphics[width=3.375in]{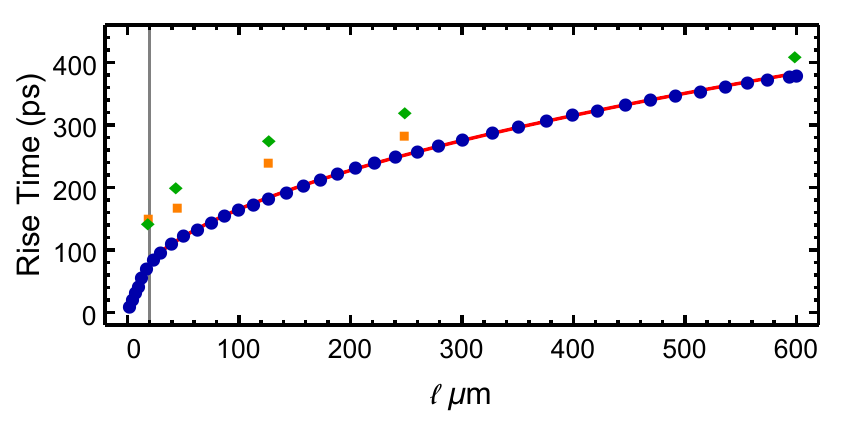}
\caption{Our prediction $t_\text{rise} \propto \sqrt{\ell}$ (red curve) compared with theoretical predictions from Smirnov \textit{et al.} \cite{smirnov2016} (blue dots), alongside their experimental results for NbN on Si/$\text{Si}_3\text{N}_4$ (green diamonds) and Si/$\text{SiO}_2$ (orange squares).  We fit their theoretical predictions for values $>$ 20 $\mu$m (gray vertical line).}
\label{smirnovres}
\end{figure}

More detailed information may be gleaned from Eq.~\ref{tch} by assuming a specific functional form for $\tilde{v}(\tilde{I})$.  We follow Kerman \textit{et al.} \cite{kerman2009electrothermal} by using an approximate solution to the phase front velocity originally derived by Broom and Rhoderick \cite{broom1960}, which in physical, unscaled units is 
\begin{equation}
v(I)=\sqrt{2}v_o\frac{I^2/I_\text{ss}^2-1}{\sqrt{I^2/I_\text{ss}^2-1/2}}.
\label{kermanvelocity}
\end{equation}
Here, $v_o=\sqrt{\alpha\kappa/d}/c$ where $\kappa$, $c$, and $d$ are the thermal conductivity, specific heat per unit volume, and thickness of the nanowire, respectively, and $\alpha$ is the heat conductivity coefficient for cooling to the substrate. Note that $v(I_\text{ss})=0$ from Eq.~\ref{kermanvelocity}, as discussed qualitatively above.  An important quantity in studying these problems is the Stekly parameter, which describes the relative magnitudes of Joule heating and cooling to the substrate, $s=2(I_{sw}/I_\text{ss})^2$ \cite{gurevich1987,kerman2009electrothermal}.  For most SNSPDs, $s$ is large, as it is for our detector with $s=$300.  A device is usually biased near $I_\text{sw}$, therefore, over most of the rise time, Eq.~\ref{kermanvelocity} may be approximated as the linear relation $v(I)=\sqrt{2}v_o I/I_\text{ss}$, and importantly $v_b=\sqrt{2} v_o I_b/I_\text{ss}$.  Under these assumptions and using Eq.~\ref{tch}, we obtain the additional scaling relation $t_\text{ch}\propto 1/\sqrt{I_b}$.  We test this prediction against our experimental observations below.

Taking this linear thermal model, Eqs.~\ref{nearuni2f} and \ref{nearuni1f} may be solved in terms of elementary functions given by

\begin{eqnarray}
\tilde{I}(\tilde{t})&=&\mathrm{sech}^2(\tilde{t}), \label{linearI} \\
\tilde{R}(\tilde{t})&=&\mathrm{tanh}(\tilde{t}), \label{linearR} \hspace{.3in} \tilde{t} \geq 0,
\end{eqnarray}
where we continue to assume that the term proportional to $1/\tilde{\tau}_r$ in Eq.~\ref{nearuni1} can be ignored.  For comparing to experiment, it is useful to restate Eq. \ref{linearI} in terms of the readout signal $\tilde{I}_s=\mathrm{tanh}^2(\tilde{t})$, and we have done so below.  Note this form of $\tilde{v}$ does not allow for detector reset and hence only describes the SNSPD turn-on dynamics.

We find that the time at which $\tilde{I}_s$ ($\tilde{R}$) reach a value of 1/2 is given by $\tilde{t}_\text{1/2}\sim 0.881$ ($\sim 0.549$), consistent with the scaling behavior discussed above for a generic thermal model.  To directly compare this model to experiment, only a knowledge of $I_b$ and two scale parameters is needed.  (See Appendices F and G for more information on extracting scale parameters.) Therefore, for large $\tau_r$ and $s$, Eqs.~\ref{linearI} and \ref{linearR} constitute a universal model for SNSPD turn-on dynamics

\begin{figure}[htb]
\centering
\includegraphics[width=3.375in]{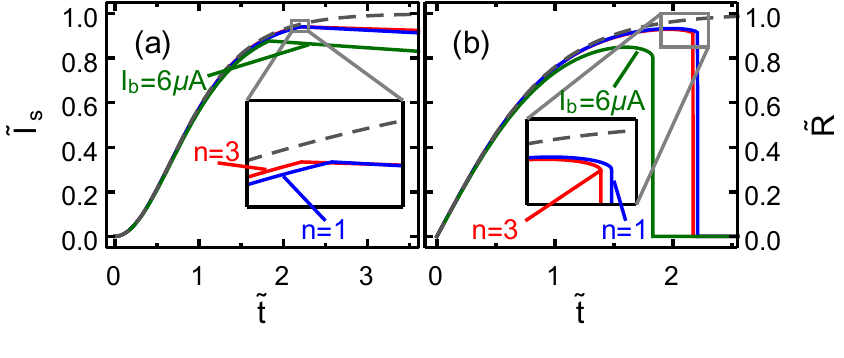}
\caption{Comparison of the exact (Eqs.~\ref{nearuni2} and \ref{nearuni1}) and universal model solutions (Eqs.~\ref{linearI} and \ref{linearR}, dashed line)  for current (a) and resistance (b).  Here, the parameters for the exact model are: (blue line) $n=1$, $I_b=12.5$ $\mu$A; (red line) $n=3$, $I_b=12.5$ $\mu$A, and (green line) $n=1$, $I_b=6$ $\mu$A.  For all curves, $R_L=50$ $\Omega$, $L_k=0.824$ $\mu$H, $v_0=60$ pm/ns, $R_\text{sq}=461$ $\Omega$, $w=70$ nm, and $I_\text{ss}=1.04$ $\mu$A.}
\label{comparison}
\end{figure}

We compare the predictions of the exact model of Ref. \cite{kerman2009electrothermal} using Eqs.~\ref{nearuni2}, \ref{nearuni1}, and \ref{kermanvelocity} without approximation, and our universal model given by Eqs.~\ref{linearI} and \ref{linearR} in Fig. \ref{comparison}.  Here, we show the temporal evolution of $\tilde{I}_s=1-\tilde{I}$ and $\tilde{R}$ for three different operating conditions: 1) one- and 2) three-photon detection events for a typical value of $I_\text{ss} \ll I_b \lesssim I_\text{sw}$; and 3) a one-photon event at $I_b \sim 6 I_\text{ss}$, which is a value smaller than that used in typical experiments.  For all cases, the universal model agrees well with the exact model up to $\tilde{t} \sim 2$.  Beyond this time, a distinct change in slope in the curves for $\tilde{I}$ predicted by the exact model appears due to the detector returning to the superconducting state ($\tilde{R}$ jumps abruptly to zero). For all curves up to $\tilde{t}=1$, the difference between the exact and universal models is $<3\%$ for $\tilde{I}$ and $<4\%$ for $\tilde{R}$, and the small disagreement is greatest for lower $I_b$, as expected.  Thus, the universal model is an excellent tool for understanding typical SNSPD turn-on dynamics. 

\begin{figure}[htb]
\centering
\includegraphics[width=3.375in]{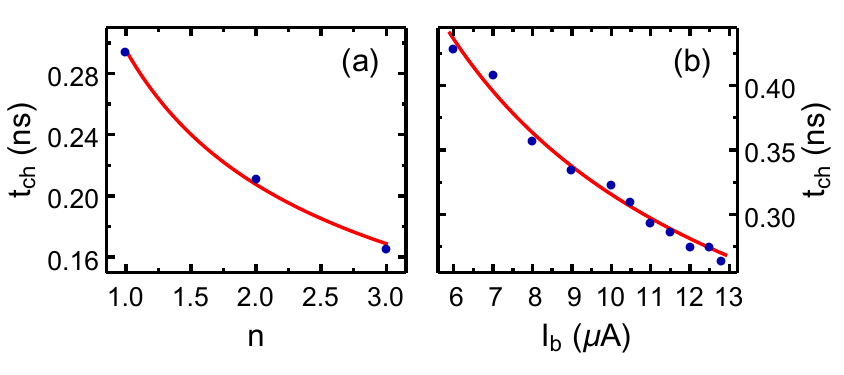}
\caption{Experimental values of $t_\text{ch}$ (blue dots) extracted by fitting with our linear model vs. (a) $n$ with $I_b$=11 $\mu$A and (b) $I_b$ with $n=1$, as well as their fits (red curves) to the model $t_\text{ch} = A/n^{a}$ or $t_\text{ch} = B/(I_b)^b$ respectively.}
\label{fits}
\end{figure}

To further explore the utility of our universal model, we use it to fit to experimental data to determine $t_\text{ch}$ for our detector.  We record $I_s$  for various values of $n$ with $I_b \sim I_\text{sw}$, and for various values of $I_b$ with $n=1$ (see Appendix F for more information).  We then fit the data using the dimensional form of Eq.~\ref{linearI} to find $t_\text{ch}$ for each data set.  The results are shown in Fig. \ref{fits}.  Considering only the photon-number data with $I_b= 10.6 I_\text{ss}$, we expect $t_\text{ch} = A/n^{a}$ with $a=0.5$.  We find $A=295 \pm 5~\text{ps}$ and $a=0.51 \pm 0.03$.  Therefore, the scaling is consistent with our universally predicted scaling.  From $A$ and other independently-measured parameters (see Supporting Materials), we find $v_0 = 96 \pm 3 ~\text{m/s}$. This value of $v_0$ is smaller than, but of-the-order-of that found for NbN-based SNSPDs \cite{berggren2018}, and is consistent with the slower turn-on and turn-off dynamics of SNSPDs based on the amorphous superconducting thin film considered here.

In contrast, considering only the bias-current data with $n=1$, we expect $t_\text{ch} = B/(I_b)^b$ with $b=0.5$; however, we find $B=1.35 \pm 0.08~\chem{ps\cdot A^{1/2}}$ and $b=0.63 \pm 0.03$.  From $B$, we predict $v_0 = 50 \pm 5~\text{m/s}$. We also simultaneously fit all data using $t_\text{ch} = C/n^c(I_b)^d$ and find $C=1.36 \pm 0.07 ~\text{ps}\cdot\chem{A^{1/2}}$, $c=0.52 \pm 0.03$, $d=0.63 \pm 0.02$, and $v_0=50 \pm 5 ~\text{m/s}$.  In both cases, the dependence of $I_b$ on $t_\text{ch}$ is inconsistent with our predicted value.  The inconsistency may be due to some physics not captured in the model proposed in Ref.~\cite{kerman2009electrothermal}.  It may also be due to high measured dark counts for higher $I_b$ measured in this detector, which may skew $t_\text{ch}$ to lower values at higher $I_b$ (see Appendix G). Finally, it may be related to distortions in the waveform caused by the amplifier as discussed above.

In conclusion, we derive a universal model for the turn-on dynamics of SNSPDs that identifies characteristic time and resistance scales, which is used to predict the observed detector behavior.  Even though there are many seemingly independent device parameters, they contribute to $t_\text{ch}$ and $R_\text{ch}$ in a highly dependent manner.  Most importantly, this model explains the multi-photon resolution observed recently in SNSPDs.  Additionally, we make further predictions on the effect of $\ell$ and $I_b$ on detector rise times and find good agreement, although our results for the latter suggest that more corrections might be needed for the model from Ref. \cite{kerman2009electrothermal}.  These observations should greatly advance our understanding on non-equilibrium dynamics of thin superconducting films exposed to light.

We gratefully acknowledge discussions of this work with Karl Berggren and Sae Woo Nam, and the financial support of the Office of Naval Research Multidisciplinary University Research Initiative program on Wavelength-Agile Quantum Key Distribution in a Marine Environment (grant \#N00014-13-1-0627) and the NASA program on Superdense Teleportation (grant \#NNX13AP35A).


\section*{Appendix A: Dependence of Kinetic Inductance on Device Current}
The kinetic inductance $L_k$ of a superconducting nanowire is a function of the current density in the nanowire. Clem and Kogan \cite{clemandkogan} have shown theoretically that near the superconductor's depairing current, $L_k$ can vary by a large ($>1/2$) fraction from its low current value. Santavicca \textit{et al.} \cite{santavicca} found experimentally that variations $\le 10\%$ are typical over the entire range of current values for which their nanowires remained superconducting. Similarly, we measured a variation of 8\% in a WSi device \cite{nicolich2017SPW}. In integrating Eq.~\ref{nearuni2f} and \ref{nearuni1f}, we assume that $L_k$ is constant. This should lead to a small error in pulse shape predictions for the initial part of the detector signal when the nanowire current is large and decreasing and its kinetic inductance is changing. This effect is below our present experimental sensitivity. 

By contrast, the scaling relations with photon number and nanowire length are entirely unaffected by  any kinetic inductance dependence on current.  To see this, we begin by defining the normal state resistance and kinetic inductance per unit length of nanowire, $\mathcal{R}=R_{max}/ \ell$ and $\mathcal{L}_k(I)=L_k (I)/ \ell$, respectively. Here  the kinetic inductance is allowed to depend on the detector current. The model Eqs. \ref{genericthermEQ} and  \ref{elecEQ} may be rewritten as
\begin{eqnarray}
\label{genericthermEQb}\frac{dR_\text{hs}(t)}{dt} = 2 \mathcal{R} v(I_\text{det}(t)) \hspace{.2in} R_\text{hs} \geq 0, \\
\ell \mathcal{L}_k(I_\text{det}(t))\frac{dI_\text{det}(t)}{dt}+n R_\text{hs}(t)I_\text{det}(t) = 0. \label{elecEQb} 
\end{eqnarray}
Consistent with the above treatment, the RHS of Eq. \ref{elecEQb} is set equal to 0. We define scaled variables $\tau$, $r$, and $i$ to arrive at the scaling relations
\begin{eqnarray}
t=\sqrt{\frac{2n}{\ell}}\,\tau, \\
R_\text{hs}(t)=\sqrt{2n\ell} \, r(\tau), \\
I_\text{det}(t)=i(\tau).
\label{scaledvariables}
\end{eqnarray}
In the scaled variables, Eqs. \ref{genericthermEQb} and  \ref{elecEQb} become
\begin{eqnarray}
\label{genericthermEQbs}\frac{dr(\tau)}{d\tau} =  \mathcal{R} v(i(\tau)) \hspace{.2in} 0\leq R_\text{hs}, \\
 \mathcal{L}_k(i(\tau))\frac{di(\tau)}{d\tau}+ r(\tau)i(\tau) = 0. \label{elecEQbs} 
\end{eqnarray}
The initial conditions for this system are $i(\tau = 0)=I_b$ and $r(\tau=0)=0$. The functions $\mathcal{L}_k(i)$ and $v(i)$ are intrinsic functions of the nanowire. Thus, for a specific bias current and nanowire, pulses with different numbers of photons or from detectors of different lengths are described by exactly the same scaled system, even if the kinetic inductance of the detector depends on current. The solution for this scaled system $i(\tau)$, $r(\tau)$ is unique and scaling relations for different risetimes for pulses of different photon numbers or from detectors with different lengths are found by expressing the scaled solutions in terms of the physical variables via  Eqs. \ref{scaledvariables}. For example, for two detectors $A$ and $B$,
\begin{equation}
\begin{split}
I^A_\text{det}\left(\sqrt{\frac{n_A}{\ell_A}}\,t\right)=i(\tau)=I^B_\text{det}\left(\sqrt{\frac{n_B}{\ell_B}}\,t\right) \\
 \Rightarrow \quad I^A_\text{det}\left(\sqrt{\frac{n_A\,\ell_B }{n_B\,\ell_A}}\,t\right)=I^B_\text{det}(t).
\label{scaledvariablesresults}
\end{split}
\end{equation}
Consequently, the functional shapes of the rising-edges of their pulses is mathematically similar in the sense of Eq.~\ref{unicurve} and Fig. \ref{n123}. 

\section*{Appendix B: Experimental Procedure}

To test the theoretical predictions described in the main paper, we perform experiments to measure the rise-time of the photon-detection waveforms as a function of $n$ as well as $I_\text{bias}$. The detector is a single-pixel meander made of a proprietary amorphous superconducting material from Quantum Opus \cite{QO}. We operate the detector at $850\,$mK on the coldfinger of a $^{4}$He sorption refrigerator made by Chase Research Cryogenics \cite{chase}, which itself is mounted on the cold finger of a closed-cycle $^4$He refrigerator housed in a custom-built cryostat.

To resolve changes in the rise-time of the electrical waveform, we use a low-noise, high-bandwidth read-out circuit, which is shown in Fig. \ref{schem}, together with the lump-element depiction of the SNSPD.  The  cryogenic pre-amplifier is model CITLF3 from Cosmic Microwave Technologies, with a specified analog bandwidth of 10-2000 MHz, a noise temperature of $4$\,K, and a gain $>30$\,dB.  An additional amplifier (Mini-Circuits ZFL1000-LN, specified bandwidth 0.1-1000 MHz) at room temperature boosts the signal well above the noise-floor of the oscilloscope we use to collect the waveforms (Agilent Infiium 80404B, $8$\,GHz analog bandwidth, 40 Gsamples/s). Note that we measured these bandwidths and found the specifications to be conservative.  As a result, in Appendix G we use the measured cutoff frequencies.

To mitigate current back-action that affects AC-coupled SNSPD readout circuits~\cite{kerman2013readout}, we use a passive cross-over network at the input to the CITLF3. This network provides a DC path-to-ground that prevents charging of the input capacitor to the amplifier without degrading the fast rise-time of the waveform~\cite{cahall2018scalable}. Waveforms are recorded at different values for $I_\text{bias}$ and $n$ using the oscilloscope.

{
\begin{figure}
\includegraphics[width=3.375in]{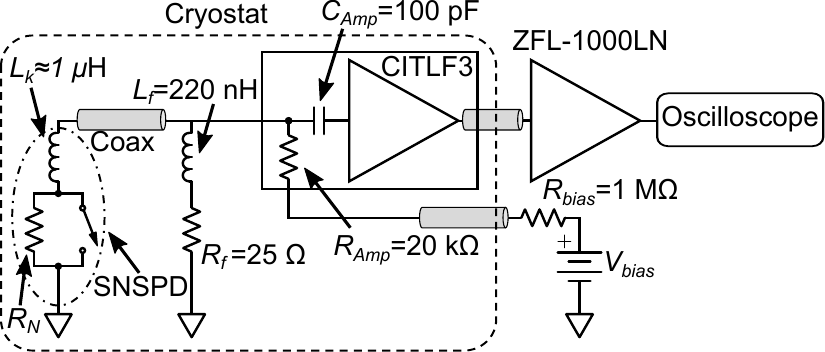}
\caption{Schematic of the experimental setup used for collecting the waveform data. The SNSPD is shown as the typical lumped-element depiction with a kinetic inductance $L_\text{k}$ and a hot-spot resistance $R_\text{N}$ that is a function of the absorbed photon number $n$.}
\label{schem}
\end{figure}
}

The source consists of a distributed feedback laser (Fitel F0L15DCWC-A82-19340-B) operating at $1550\,$ nm that is intensity modulated via electro-optic modulators (EOSpace). The repetition rate and width of the modulation signal is controlled by a field-programmable gate array (FPGA; Altera Stratix V 5SGXEA7N2F40C2), shown in Fig. \ref{source_schem}. Two sequential modulators are used to increase the overall extinction ratio between the ``ON" and ``OFF" states. After creating the pulse with the modulators, it is attenuated to the desired mean photon number per pulse with a variable optical attenuator before traveling to the detector. In the data presented in this paper, the width of the modulated optical pulse is $\sim80\,$ps, the repetition rate is 610\,kHz, and the mean photon number per pulse is $\sim$ 1.26.

{
\begin{figure}
\includegraphics[width=3.375in]{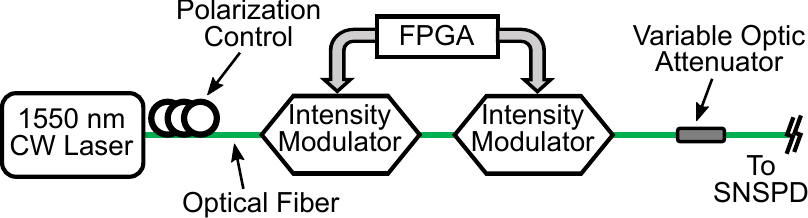}
\caption{Schematic of the source setup used in generating multi-photon wavepackets. }
\label{source_schem}
\end{figure}
}

\section*{Appendix C: Effect of the Finite Bandwidth of the Detection System}

We have assumed in the main text that the system gain for the SNSPD signal is flat as a function of frequency and the bandwidth is infinite. The most significant departure from this in the experimental apparatus is our system's high frequency limits. The expected decrease in system gain at high frequencies is dominated by  the room temperature amplifier, whose gain we measured to be down by 3~dB at $f_{lp}=2.4$ GHz. Accordingly, we model the system gain's high frequency roll off by a single-pole low pass filter with a time constant $\tau=2\pi/f_{lp}$. Here, $R$ and $C$ are chosen to reproduce the amplifier's 3~dB corner frequency. For this model, the ``actual'' signal, $v_\text{actual}(t)$ that would be observed for an infinite bandwidth system is expressed in terms of the measured, filtered signal $v_\text{measured}(t)$,
\begin{equation}
    v_\text{actual}(t)=\tau \frac{d v_\text{measured}(t)}{dt}+v_\text{measured}(t).
    \label{filterexpression}
\end{equation}

The derivative signal in Fig. \ref{n123} shows this effect most clearly.  Peak  heights are inversely proportional to the pulse risetimes, and the finite bandwidth of the system means that the measured heights of these curves will be somewhat smaller than the ``actual" heights that result from amplifiers with infinite bandwidth. We mathematically model the ``measured" curves as Gaussians whose full width at half maximum are taken from the data shown in the figure. Using Eq. \ref{filterexpression}, we predict that for events of 1, 2, and 3, photons, the ``actual" peaks were reduced by 6\%, 11\%, and 15\%, respectively to give those measured values in the figure. Equivalently, the actual lifetimes are expected to be shorter than the measured risetimes by the same amounts, according to this model of the amplifier system's frequency dependence. This size of an effect is right at the edge of our experimental sensitivity given uncertainties in the exact shape of the pulse risetime distributions and we have not identified it in our data. Future, higher precision work and, especially, an extension of this technique to higher photon-number events, will require care on this score.

\section*{Appendix D: Comparison of Raw Waveforms to Rescaled Waveforms for Different $n$}
\label{stats}
To determine whether there is a statistical significance between the waveforms shown in Fig. 2 of the main text, we performed a one-way ANOVA \cite{statbook} on the derivative waveforms at $t=0$.  Typical one-way ANOVA results are reported using an $F$ statistic, which represents the ratio of the variance between groups to the variance within groups as a function of degrees-of-freedom between the groups and total degrees-of-freedom, and a $p$-value representing significance level, where $p<0.05$ is considered statistically significant.  For the raw waveforms, the results are F(2,12)$=140,p=\num{5.6e-9}$, which shows a statistically significant difference.  Post hoc comparisons using the Tukey HSD test show a significant difference between the $n=1$ and $n=2$, $n=2$ and $n=3$, and $n=1$ and $n=3$ groups.  In contrast, for the rescaled waveforms, we find F(2,12)$=1.7,p=0.22$, indicating no statistical significance between the waveforms at $t=0$.

\section*{Appendix E: Digitization and Fitting of Smirnov Data}
\label{smirnovdat}
The data used to compare the Smirnov \textit{et al.} two-temperature predictions with our theory is digitized from \cite{smirnov2016} using the software Plot Digitizer \cite{plotdig} and is shown in Fig. \ref{smirnovres} alongside their experimental results.  As described in the main text, our theory agrees well with theirs with a $\chi_r^2\sim1$ and agrees qualitatively with their experimental findings.  When calculating the $\chi_r^2$, we assume that the dominant error is the $\sim$1.6 ps error introduced by digitizing.  The digitization program works by allowing the user to set the axes and hand-select data points to record their values.  We arrived at this digitization error by selecting the same data point multiple times.

\section*{Appendix F: Extracting $t_\text{ch}$ from Experimental Pulses}
\label{extch}
\begin{figure}
    \centering
    \includegraphics[width=3.375in]{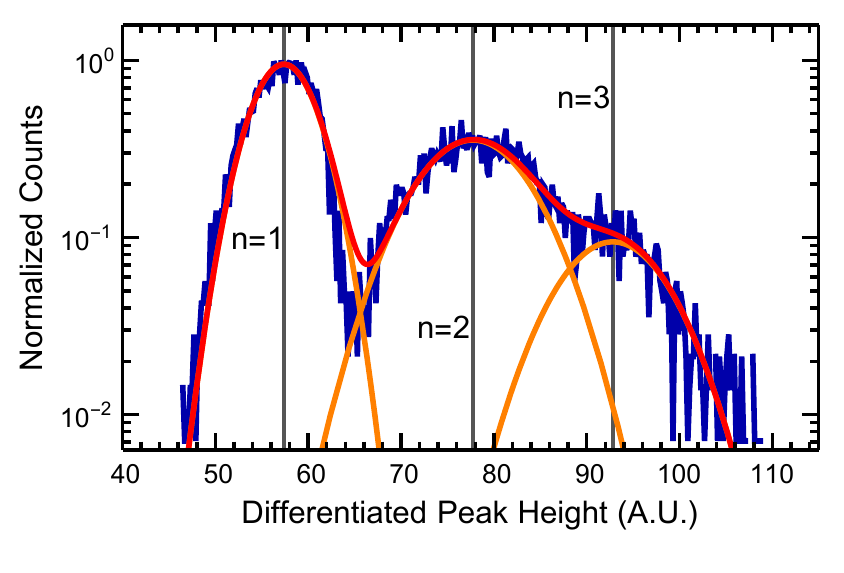}
    \caption{Histogram of differentiated peak heights (blue) for $I_b$=11 $\mu$A.  We fit a sum of Gaussian functions (red) of the form $\sum_{n=1}^3 a_n e^{-((x-b_n)/c_n)^2}$ to the distribution.  Resulting individual Gaussians (orange curves) and their centers (gray vertical lines) are also shown.}
    \label{histofit}
\end{figure}

\begin{figure}
    \centering
    \includegraphics[width=3.375in]{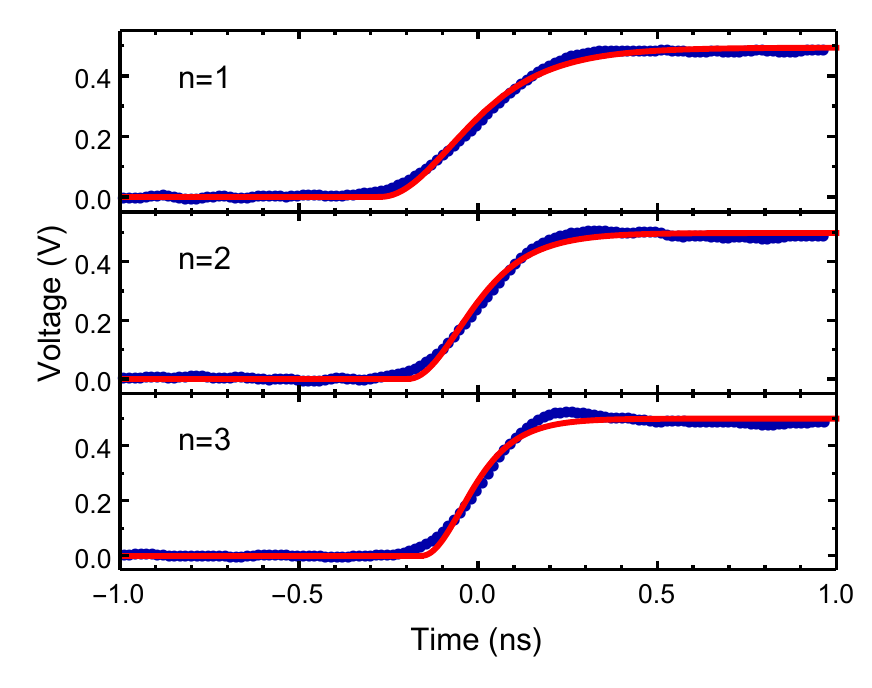}
    \caption{Universal model fits (red solid line) for $n=$1, 2, and 3 (blue points).}
    \label{nfits}
\end{figure}

To arrive at values for $t_\text{ch}$ for varying $n$, we follow \cite{cahall2017heckyes} and make a histogram of the maximum values of the derivatives of readout pulses as shown in Fig. \ref{histofit}. We fit a sum of Gaussians to the distribution, finding $\chi_r^2\sim2.2$, and then use each Gaussian's center to determine representative pulses for $n$=1, 2, and 3.  We then fit these representative pulses using our universal model (Eq. \ref{linearI}) to determine $t_\text{ch}$, as shown in Fig. \ref{nfits}.  The resulting values are given in Table \ref{tchtable}.  Note that with a value of $t_\text{ch}$ extracted in this manner for a pulse with known $n$, combined with a measurement of $L_k$ as described below, it is possible to extract $R_\text{ch}$ as well.

\begin{figure*}
    \centering
    \includegraphics[width=6.5in]{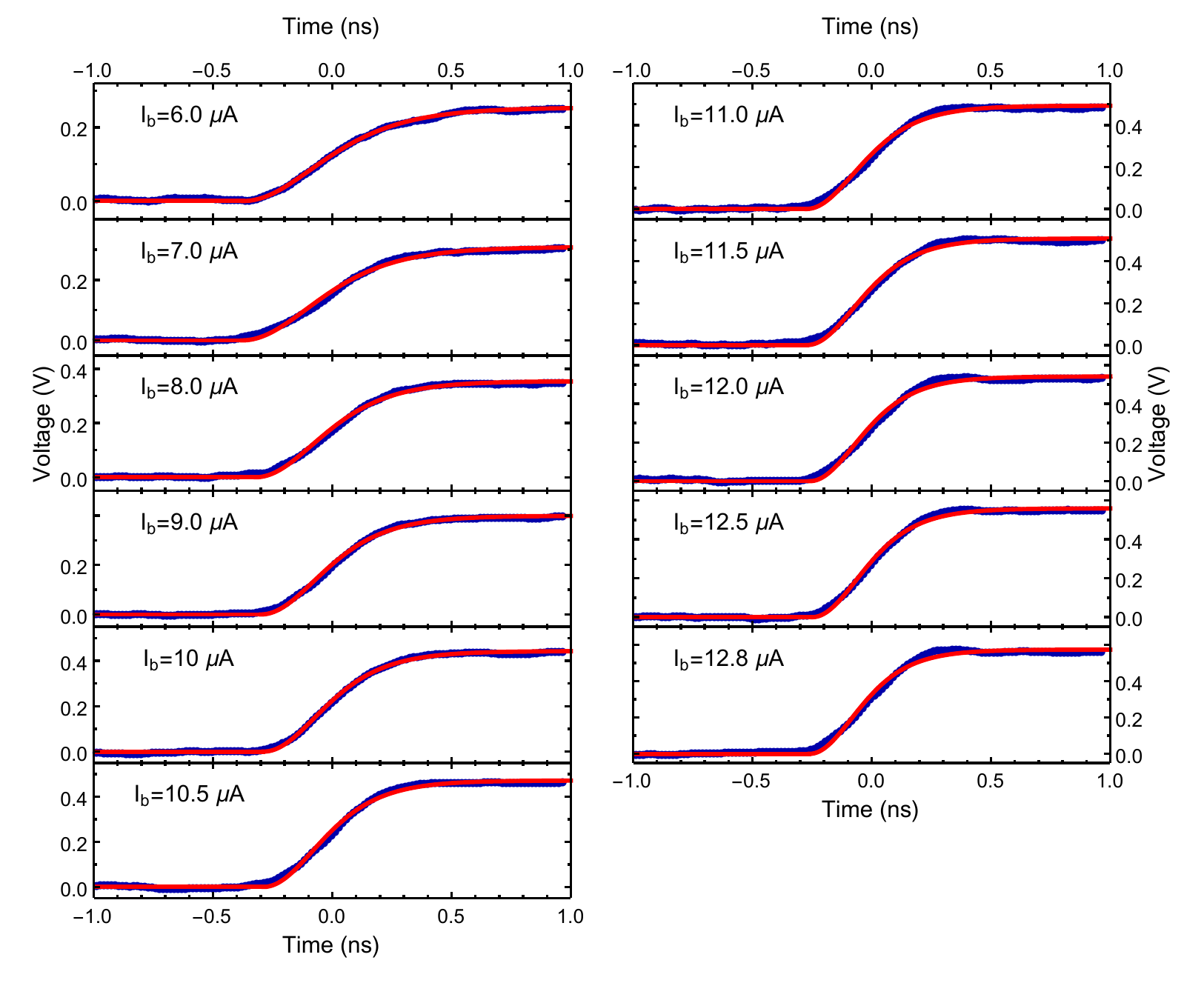}
    \caption{Universal model fits (red solid line) for different $I_b$ (blue points).}
    \label{ibfits}
\end{figure*}

For varying $I_b$, we use the same procedure and focus only on the $n=1$ peak for each value of $I_b$.  Fits to the waveforms can be found in Fig. \ref{ibfits} and values of $t_\text{ch}$ are given in Table \ref{tchtable}.

The $\chi_r^2$ for these fits ranged from 1.1 to 11.5 for $I_b=6$ to $12.8~\mu A$, and 6.6 to 14.1 for $n=1$ to 3.  We hypothesize that the increase of $\chi_r^2$ is due somewhat to the limited bandwidth of and ringing behavior observed in our amplifier.  The rising edge for $n=3$ and $I_b$=12.8 $\mu$A is somewhat faster than the amplifier rise time, and therefore the fit slightly underestimates the actual data when the pulse first begins to rise at $\sim-0.15$ ns.  Similarly, when the pulse begins to round off at its maximum value at $\sim0.2$ ns, the amplifier rings, again causing the fit to somewhat undershoot the actual data.  This effect should be more pronounced for higher $n$ and $I_b$ where the rise time is shortest, and this is reflected in the higher $\chi_r^2$ for these fits. 

\begin{table}
\begin{tabular}{ |p{0.4cm}|p{1.2cm}||p{2.0cm}|  }
 \hline
 $n$ & $I_b$ ($\mu$A) & $t_\text{ch}$ (ns)\\
 \hline
 1 & 12.8 & 0.264 $\pm$ 0.007\\
 1 & 12.5 & 0.275 $\pm$ 0.005\\
 1 & 12.0 & 0.276 $\pm$ 0.007\\
 1 & 11.5 & 0.287 $\pm$ 0.006\\
 1 & 11.0 & 0.294 $\pm$ 0.007\\
 2 & 11.0 & 0.212 $\pm$ 0.006\\
 3 & 11.0 & 0.166 $\pm$ 0.006\\
 1 & 10.5 & 0.310 $\pm$ 0.006\\
 1 & 10.0 & 0.323 $\pm$ 0.004\\
 1 & 9.0 & 0.335 $\pm$ 0.004\\
 1 & 8.0 & 0.358 $\pm$ 0.006\\
 1 & 7.0 & 0.409 $\pm$ 0.008\\
 1 & 6.0 & 0.429 $\pm$ 0.007\\
 \hline
\end{tabular}\\
\caption{Values of $t_\text{ch}$ found by fitting experimental waveforms to the universal model solution.}
\label{tchtable}
\end{table}

\section*{Appendix G: Determining Detector Parameters}
\label{detparams}
\subsection{Steady State Current}
\begin{figure}
    \centering
    \includegraphics[width=3.375in]{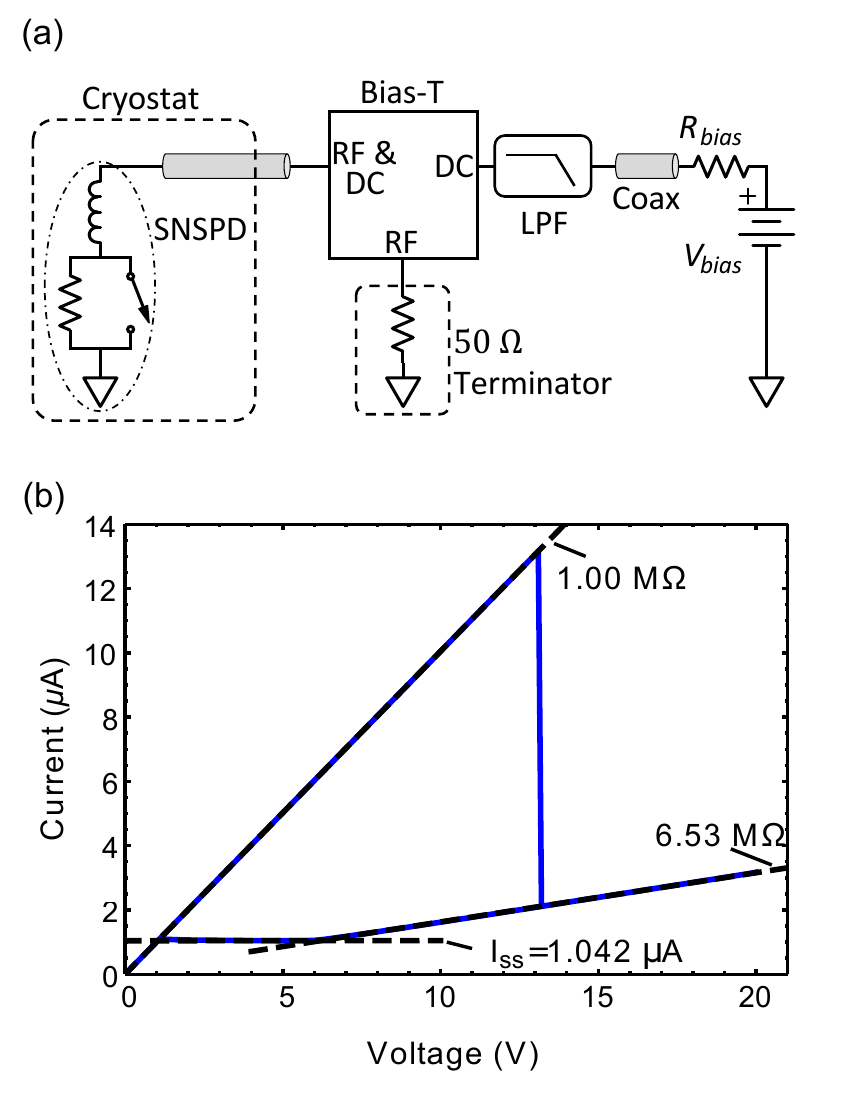}
    \caption{Measurement schematic (a) and resulting IV curve (b) for the Quantum Opus detector. There is a 1 M$\Omega$ series resistor $R_\text{bias}$ between the voltage source and the detector, resulting in a slope of 1 M$\Omega$ during the superconducting portion of the IV curve.  The normal resistance is about 6.53 M$\Omega$ including $R_\text{bias}$. The bias-T is a Mini-Circuits ZFBT-6GW and the low pass filter (LPF) is a Mini-Circuits SLP-1.9+.}
    \label{ivcurve}
\end{figure}

The steady-state current $I_\text{ss}$ is defined as the current when the Joule heating and cooling to the substrate are balanced such that $v(I_\text{ss})=0$.  We determine $I_\text{ss}$ by performing a DC current-voltage (IV) measurement as shown in Fig. \ref{ivcurve}.  We start with a high voltage such that the whole meander is in the normal state.  We gradually lower the voltage, and observe a plateau region starting at around 6 V and ending around 1 V.  In this region, sections of the nanowire are beginning to switch into the superconducting state, causing the resistance to drop.  In turn, the current maintains a constant value, $I_\text{ss}$.  We fit a line to the IV curve from 4-6V and find the vertical offset is $I_\text{ss}=1.042 \pm 0.005~\mu \text{A}$.

\subsection{Kinetic Inductance}
\begin{figure}
    \centering
    \includegraphics[width=3.375in]{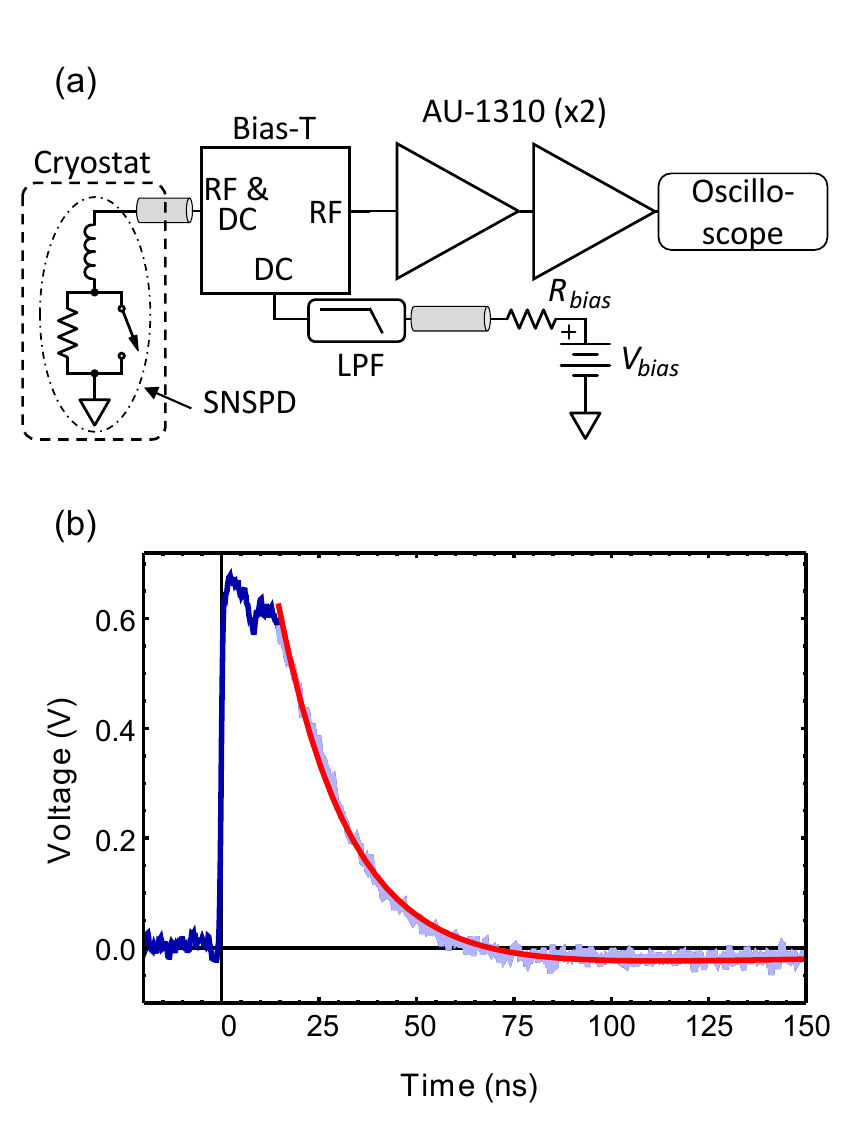}
    \caption{Measurement schematic (a) and resulting SNSPD pulse (b).  We fit Eq. \ref{lkmodel} (red) to the falling edge (light blue region) and extract $L_k$. The low pass filter (LPF) is a Mini-Circuits SLP-1.9+.}
    \label{lkcurve}
\end{figure}

To determine $L_k$, we model the electrical properties of the SNSPD during the falling edge of the readout pulse and fit to experimental waveforms.  We place the detector in parallel with two room-temperature amplifiers using a bias-T (Mini-Circuits ZFBT-6GW). Typically, it is assumed that the fall time for $I_s$ is given by $\tau_r=L_k/R_L$.  In this case, we modify this assumption to also allow for timing variations caused by the AC-coupling capacitor.  We use the model
\begin{equation}
\begin{split}
    I_s(t)=e^{(R_L/2 L_k)(t-t_o)}(a e^{\sqrt{(R_L/2 L_k)^2-1/(L_k C)}(t-t_o)} \\
    +b e^{-\sqrt{(R_L/2 L_k)^2-1/(L_k C)}(t-t_o)})
    \label{lkmodel}
\end{split}
\end{equation}
where $R_L=$50 $\Omega$ and $a$, $b$, $t_o$, $L_k$, and $C$ are left as free parameters. ($C$ is not specified by the bias-T manufacturer, so we allow it to be a fit parameter.) We fit Eq. \ref{lkmodel} to the pulse shown in Fig. \ref{lkcurve} from $t\sim15$ to $t=250$ ns.  We find $a=-0.059 \pm 0.004~\text{V}$, $b=1.53 \pm 0.01~\text{V}$, $t_o=-1.2722 \pm 0.0004~\text{ns}$, $C=3.2 \pm 0.2~\text{nF}$, and $L_k= 824 \pm 4~\text{nH}$.  The fit has $\chi_r^2\sim1.6$.

\subsection{Dark Counts}
\begin{figure}
    \centering
    \includegraphics[width=3.375in]{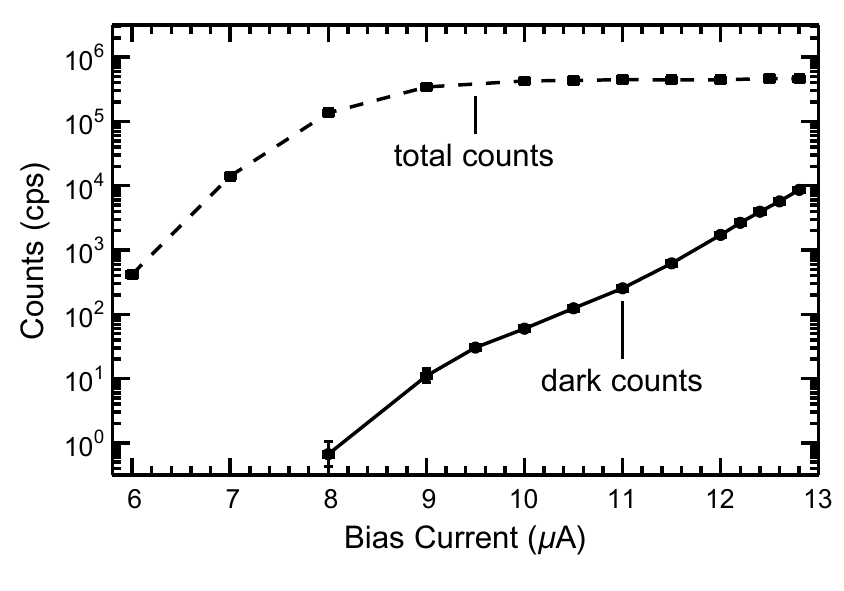}
    \caption{The detector dark count rate (points with solid line) varies from a few counts per second (cps) for low $I_b$, up to $>8,000$ cps for high $I_b$.  The total count rate (squares with dotted line) is also shown.  (Note measured results are given by plot points -- the lines are just guides for the eye.)}
    \label{darkrate}
\end{figure}

The dark count rate (DCR) as well as the total count rate (TCR) as a function of $I_b$ for the detector is shown in Fig. \ref{darkrate}. At higher $I_b$ ($\gtrsim$ 11 $\mu$A), the DCR grows to within two orders-of-magnitude of the TCR.

\begin{figure}
\centering
\includegraphics[width=3.375in]{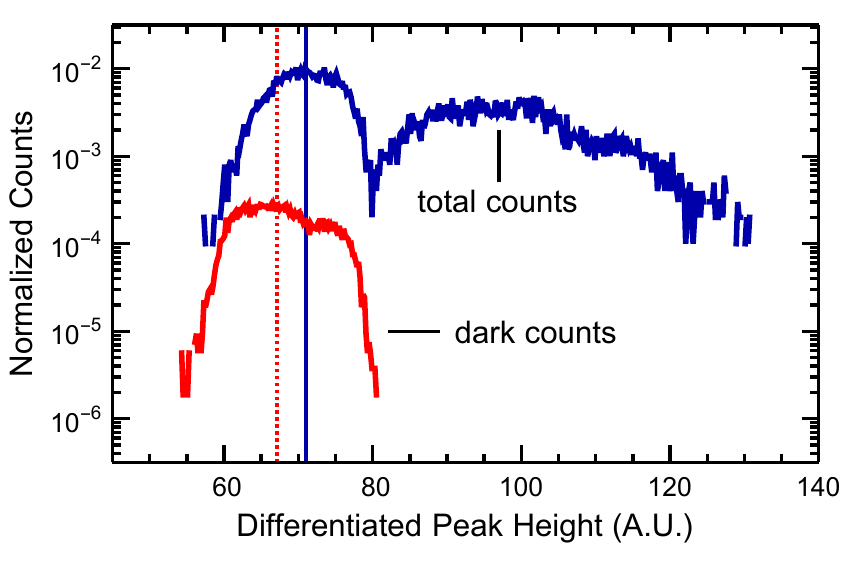}
\caption{Comparison of multi-photon absorption data (blue) with dark count rates (red) at $I_b=12.8 \mu$A. Both distributions are normalized such that the area under their curves are unity, then the dark counts are further scaled by the ratio of the DCR over the overall count rate during data collection.  Gaussians are fit to the $n=1$ peak and their centers are shown as vertical lines for both source (solid blue) and dark (dotted red) counts.}
\label{dc}
\end{figure}

Careful analysis of our dark counts reveals an interesting phenomenon, which is not accounted for by our model.  We collect dark count pulses, differentiate their rising edges, and record their maximum derivatives in the same way as above.  We assume that we do not obtain any multi-photon dark count events and so the resulting distribution represents only $n=1$ counts. We compare the resulting distribution to our multi-photon distribution and find that the $n=1$ dark counts are shifted to lower differentiated peak height (longer rise time) as shown in Fig. \ref{dc}.  A possible explanation for this shift is that the dark count photons are likely much longer wavelength than our source photons \cite{shibata2015}; however, there is no mechanism in our model to account for this proposed difference.  This is an important issue that warrants further exploration.

\FloatBarrier
%

\end{document}